\documentclass{sig-alternate} 

\usepackage{tabularx}
\usepackage{booktabs}
\usepackage{datetime}
\usepackage{subcaption}
\usepackage[ruled,vlined,linesnumberedhidden,lined,boxed,commentsnumbered]{algorithm2e}
\usepackage{hyperref}
\usepackage{paralist}
\usepackage{multirow}
\usepackage{pbox}
\usepackage{bbm}
\usepackage{arydshln}

\newtheorem{definition}{Definition}

\newfont{\mycrnotice}{ptmr8t at 7pt}
\newfont{\myconfname}{ptmri8t at 7pt}
\let\crnotice\mycrnotice%
\let\confname\myconfname%

\permission{Permission to make digital or hard copies of all or part of this work for personal or classroom use is granted without fee provided that copies are not made or distributed for profit or commercial advantage and that copies bear this notice and the full citation on the first page. Copyrights for components of this work owned by others than ACM must be honored. Abstracting with credit is permitted. To copy otherwise, or republish, to post on servers or to redistribute to lists, requires prior specific permission and/or a fee. Request permissions from Permissions@acm.org.} 
\copyrightetc{\copyright~2015 ACM. ISBN \the\acmcopyr}
\crdata{978-1-4503-3794-6/15/10\ ...\$15.00.\\
DOI: http://dx.doi.org/10.1145/2806416.2806531}

\begin{document}


\newcommand{\seq}[1]{\langle\,{#1}\,\rangle}
\newcommand{\set}[1]{\left\{\,{#1}\,\right\}}
\newcommand{\bigset}[1]{\left\{\,{#1}\,\right\}}
\newcommand{\bigtuple}[1]{\left(\,{#1}\,\right)}
\newcommand{\card}[1]{\left|\,{#1}\,\right|}
\newcommand{\tup}[1]{\left(\,{#1}\,\right)}
\newcommand{\argmin}[2]{\underset{#1}{\operatorname{arg\,min}}{\:\: #2}}
\newcommand{\stdmin}[2]{\underset{#1}{\operatorname{min}}{\:\: #2}}
\newcommand{\simplemin}[1]{\ensuremath{\underset{#1}{min}\;}} 
\newcommand{\pow}[2]{\ensuremath{#1^#2\;}}
\newcommand{\comment}[1]{}
\newcommand{\ra}[1]{\renewcommand{\arraystretch}{#1}}
\newcommand {\N}{\mathcal{N}}
\newcommand{\argmax}{\operatornamewithlimits{argmax}}

\hyphenation{Map-Reduce}
\hyphenation{opti-mi-za-tion}
\hyphenation{Wiki-pedia}
\hyphenation{sali-ence}

\title{Automated News Suggestions for Populating Wikipedia Entity Pages}

\numberofauthors{1} \author{ \alignauthor
  Besnik Fetahu$^\dagger$,  Katja Markert$^\dagger$$^\ddagger$,   Avishek Anand$^\dagger$\\
  \affaddr{
    \begin{tabular}{c@{~~~~}c}
      \@{}&\@{}\\
      $^\dagger$L3S Research Center, Leibniz University of Hannover&
      $^\ddagger$School of Computing, University of Leeds\\
      Hannover, Germany & United Kingdom  \\
    \end{tabular}
  } \affaddr{\{fetahu,markert,anand\}@L3S.de}
}

\maketitle

\begin{abstract} 

Wikipedia entity pages are a valuable source of information for direct consumption and for knowledge-base construction, update and maintenance. Facts in these entity pages are typically supported by references. Recent studies show that as much as 20\% of the references are from online news sources. However, many entity pages are incomplete even if relevant information is already available in existing news articles. Even for the already present references, there is often a delay between the news article publication time and the reference time. In this work, we therefore look at Wikipedia through the lens of news and propose a novel news-article suggestion task to improve news coverage in Wikipedia, and reduce the lag of newsworthy references. Our work finds direct application, as a precursor, to Wikipedia page generation and knowledge-base acceleration tasks that rely on relevant and high quality input sources.

We propose a two-stage supervised approach for suggesting news articles to entity pages for a given state of Wikipedia. First, we suggest news articles to Wikipedia entities (article-entity placement) relying on a rich set of features which take into account the \emph{salience} and \emph{relative authority} of entities, and the \emph{novelty} of news articles to entity pages. Second, we determine the exact section in the entity page for the input article (article-section placement) guided by class-based section templates. We perform an extensive evaluation of our approach based on ground-truth data that is extracted from external references in Wikipedia. We achieve a high precision value of up to 93\% in the \emph{article-entity} suggestion stage and upto 84\% for the \emph{article-section placement}. Finally, we compare our approach against competitive baselines and show significant improvements.
\end{abstract}

\category{H3.3}{Information Systems}{Information Storage and Retrieval}[Information Search and Retrieval]

\section{Introduction}
\label{sec:introduction}

Wikipedia is the largest source of open and collaboratively curated knowledge in the world. Introduced in 2001, it has evolved into a reference work with around 5m pages for the English Wikipedia alone. In addition, entities and event pages are updated quick\-ly via collaborative editing and all edits are encouraged to include source citations, creating a knowledge base which aims at being both timely as well as authoritative. As a result, it has become the preferred source of information consumption about entities and events\footnote{Wikipedia is one of the Top 10 viewed page sites and the top reference site according to Alexa Internet ranking \url{www.alexa.com}.}. Moreso, this knowledge is harvested and utilized in building knowledge bases like YAGO~\cite{suchanek_yago:_2007} and DBpedia~\cite{bizer_dbpedia_2009}, and used in applications like text categorization~\cite{wang_building_2008}, entity disambiguation~\cite{Hoffart:2011:RDN:2145432.2145521}, entity ranking~\cite{kaptein_entity_2010} and distant supervision~\cite{surdeanu2010,Mintz:2009}.

\begin{figure}[t!]
\centering
  \includegraphics[width=0.9\columnwidth]{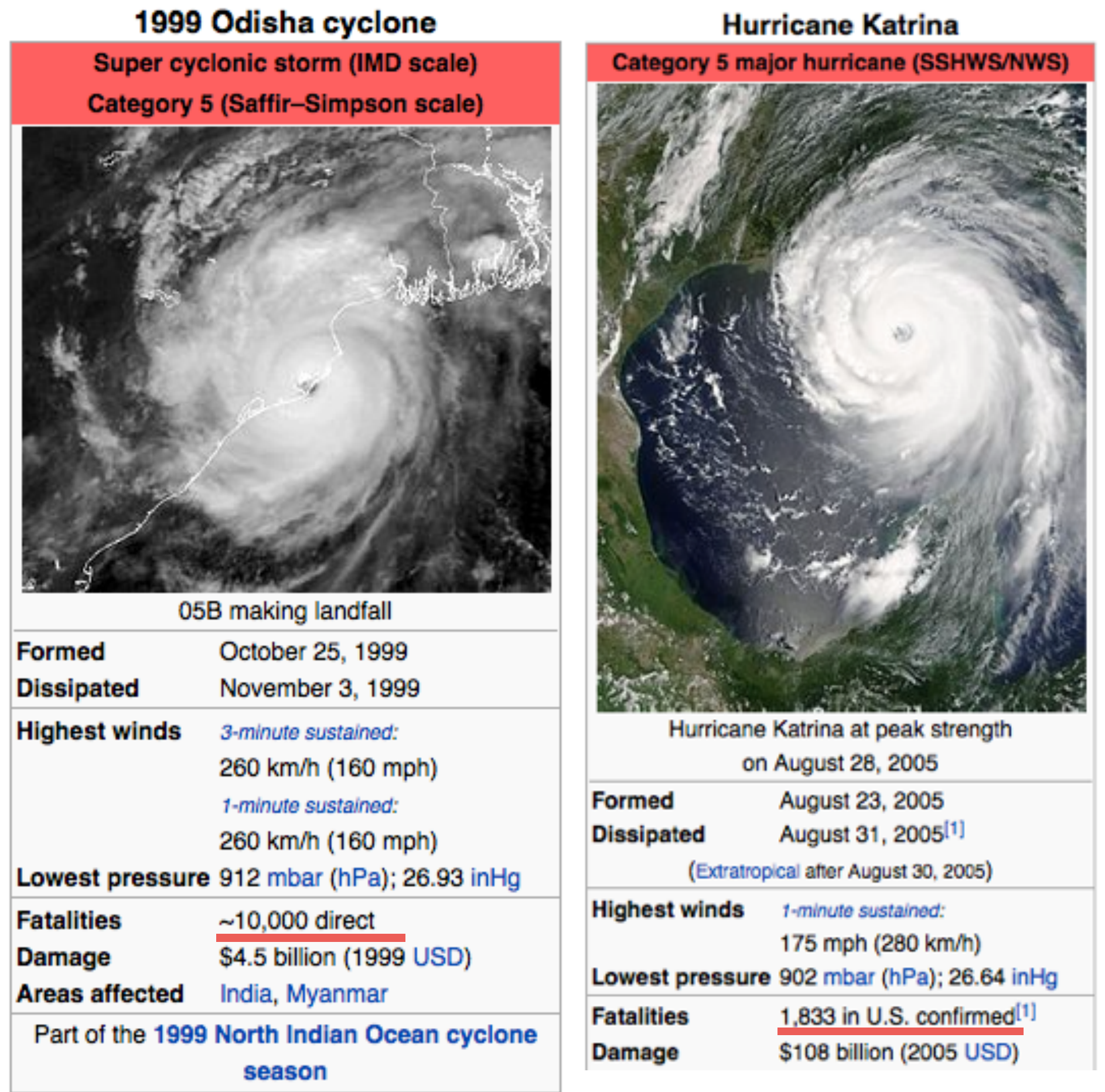}
  \caption{\small{Comparing how cyclones are reported in Wikipedia entity pages.}}
  \label{fig:cyclone}
\end{figure}

However, not all Wikipedia pages referring to entities (entity pages) are comprehensive:  relevant information can either be \emph{missing} or added with a \emph{delay}. Consider the city of \emph{New Orleans} and the state of \emph{Odisha} which were severely affected by cyclones \emph{Hurricane Katrina} and \emph{Odisha Cyclone}, respectively. While \emph{Katrina} finds extensive mention in the entity page for \emph{New Orleans}, \emph{Odisha Cyclone} which has 5 times more human casualties (cf. Figure~\ref{fig:cyclone}) is not mentioned in the page for \emph{Odisha}. Arguably \emph{Katrina} and \emph{New Orleans} are more popular entities, but \emph{Odisha Cyclone} was also reported extensively in national and international news outlets. This highlights the lack of important facts in trunk and long-tail entity pages, even in the presence of relevant sources. In addition, previous studies have shown that there is an inherent delay or lag when facts are added to entity pages~\cite{DBLP:conf/websci/FetahuAA15}.

To remedy these problems, it is important to identify information sources that contain novel and salient facts to a given entity page. However, not all information sources are equal. The online presence of major news outlets is an authoritative source due to active editorial control and their articles are also a timely container of facts. In addition, their use is in line with current Wikipedia editing practice, as is shown in~\cite{DBLP:conf/websci/FetahuAA15} that almost 20\% of current citations in all entity pages are news articles. We therefore propose \emph{news suggestion} as a novel task that enhances entity pages and reduces delay while keeping its pages authoritative. 

Existing efforts to  populate Wikipedia~\cite{DBLP:conf/acl/SauperB09} start from an entity page and then generate candidate documents about this entity using an external search engine (and then post-process them). However, such an approach lacks in (a) reproducibility since rankings vary with time with obvious bias to recent news (b) maintainability since document acquisition for each entity has to be periodically performed. To this effect, our news suggestion considers a news article as input, and determines if it is valuable for Wikipedia. Specifically, given an input news article $n$ and a state of Wikipedia, the news suggestion problem identifies the entities mentioned in $n$ whose entity pages can improve upon suggesting $n$. Most of the works on knowledge base acceleration~\cite{DBLP:conf/riao/BalogRTN13,DBLP:conf/sigir/BalogR13,DBLP:conf/eacl/DunietzG14}, or Wikipedia page generation~\cite{DBLP:conf/acl/SauperB09} rely on high quality input sources which are then utilized to extract textual facts for Wikipedia page population. In this work, we do not suggest snippets or paraphrases but rather entire articles which have a high potential importance for entity pages. These suggested news articles could be consequently used for extraction, summarization or population either manually or automatically -- all of which rely on high quality and relevant input sources. 

We identify four properties of good news recommendations: \emph{salience}, \emph{relative authority}, \emph{novelty} and \emph{placement}. First, we need to identify the most salient entities in a news article. This is done to avoid pollution of entity pages with only marginally related news. Second, we need to determine whether the news is important to the entity as only the most relevant news should be added to a precise reference work. To do this, we compute the \emph{relative authority} of all entities in the news article: we call an entity more authoritative than another if it is more popular or noteworthy in the real world. Entities with very high authority have many news items associated with them and only the most relevant of these should be included in Wikipedia whereas for entities of lower authority the threshold for inclusion of a news article will be lower. Third, a good recommendation should be able to identify \emph{novel} news by minimizing redundancy coming from multiple news articles. Finally, addition of facts is facilitated if the recommendations are fine-grained, i.e., recommendations are made on the section level rather than the page level (\emph{placement}).

\textbf{Approach and Contributions.} We propose a two-stage news suggestion approach to entity pages. In the first stage, we determine whether a news article should be suggested for an entity, based on the entity's \emph{salience} in the news article, its  \emph{relative authority} and the \emph{novelty} of the article to the entity page. The second stage takes into account the class of the entity for which the news is suggested and constructs \emph{section templates} from entities of the same class. The generation of such templates has the advantage of suggesting and expanding entity pages that do not have a complete
section structure in Wikipedia, explicitly addressing long-tail and trunk entities. Afterwards, based on the constructed template our method determines the best fit for the news article with one of the sections.

We evaluate the proposed approach on a news corpus consisting of 351,982 articles crawled from the \emph{news} external references in Wikipedia from 73,734 entity pages. Given the Wikipedia snapshot at a given year (in our case [2009-2014]), we suggest news articles that might be cited in the coming years. The existing news references in the entity pages along with their reference date act as our ground-truth to evaluate our approach. In summary, we make the following contributions.

\begin{itemize}
\itemsep0em
	\item  we propose a two-stage news suggestion approach for Wikipedia entity pages.
  \item we adopt and address the problem of determining whether a news article should be referenced to an entity considering the entity \emph{salience}, \emph{relative authority} and \emph{novelty} of the article for the entity page.
  \item we are able to place articles in a specific section of the entity page. Through \emph{section templates}, we address the problems of entities with a limited section structure by
class-based generalization i.e. we can expand entity pages with sections that come from entities of a similar class.
  \item an extensive evaluation on 351,982 news articles and 73,734 entity pages, using  their state for the years [2009-2013].
\end{itemize}

\begin{figure}[ht!]
\centering
  \includegraphics[width=0.5\textwidth]{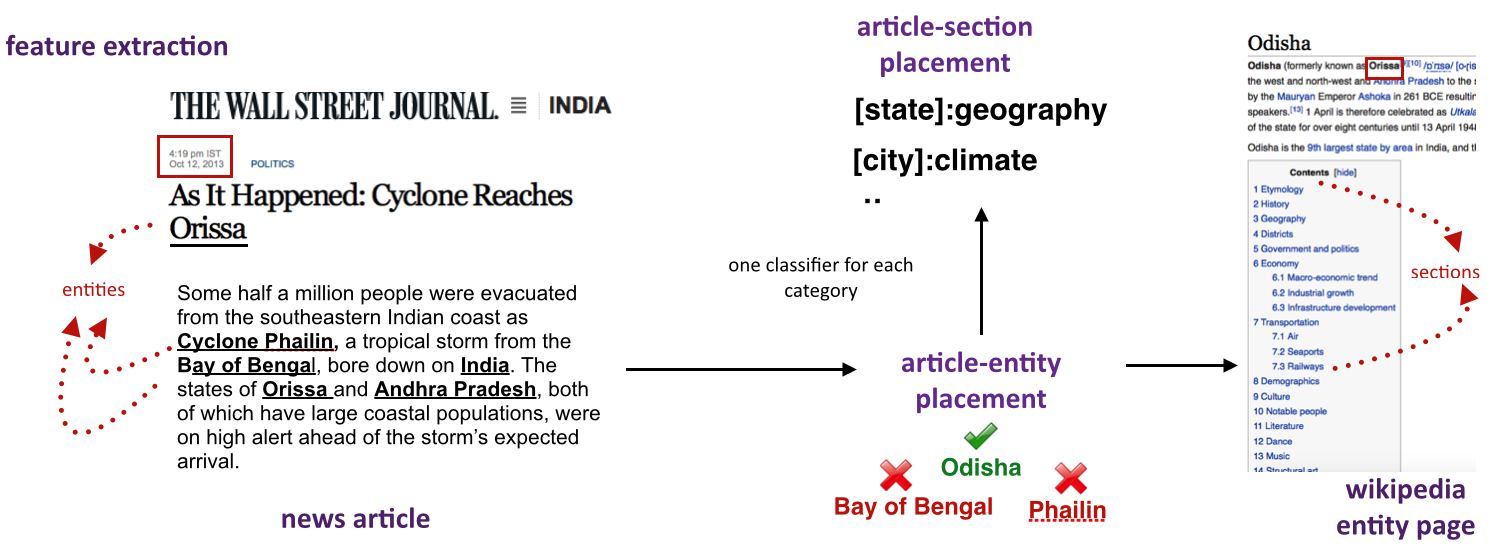}
  \caption{\small{News suggestion approach overview.}}
  \label{fig:approach}
\end{figure}

\section{Related Work}
\label{sec:related-work}

As we suggest a new problem there is no current work addressing exactly the same task. However, our task has similarities to Wikipedia page generation and knowledge base acceleration. In addition, we take inspiration from Natural Language Processing (NLP) methods for salience detection.

\textbf{Wikipedia Page Generation} is the problem of populating Wikipedia pages with content coming from external sources. Sauper and Barzilay \cite{DBLP:conf/acl/SauperB09} propose an approach for
automatically generating  whole  entity pages for specific entity classes. The approach is trained on already-populated entity pages of a given class (e.g. `\emph{Diseases}') by learning templates about the entity page structure (e.g. diseases have a \emph{treatment} section). For a new entity page, first, they extract documents via Web search using the entity title and the section title as a query, for example `\emph{Lung Cancer}'+`\emph{Treatment}'. As already discussed in the introduction, this has problems with reproducibility and maintainability. However, their main focus is on identifying the best  paragraphs extracted from the collected documents. They rank the paragraphs via an optimized supervised \emph{perceptron model} for finding the most representative paragraph that is the least similar to paragraphs in other sections. This paragraph is then included in the newly generated entity page. Taneva and Weikum~\cite{Taneva:2013:GEM:2505515.2505715} propose an approach that constructs short summaries for the long tail. The summaries are called `\emph{gems}' and the size of a `\emph{gem}' can be user defined. They focus on generating summaries that are novel and diverse. However, they do not consider any structure of entities, which is present in Wikipedia.
 
In contrast to \cite{DBLP:conf/acl/SauperB09} and \cite{Taneva:2013:GEM:2505515.2505715}, we actually focus on suggesting entire documents to Wikipedia entity pages. These are authoritative documents (news), which are highly relevant for the entity, novel for the entity and in which the entity is salient. Whereas relevance in Sauper and Barzilay is implicitly computed by web page ranking we solve that problem by looking at relative authority and salience of an entity, using the news article and entity page only. As Sauper and Barzilay concentrate on empty entity pages, the problem of novelty of their content is not an issue in their work whereas it is in our case which focuses more on updating entities. Updating entities will be more and more important the bigger an existing reference work is. Both the approaches in~\cite{DBLP:conf/acl/SauperB09} and \cite{Taneva:2013:GEM:2505515.2505715}  (finding paragraphs and summarization) could then be used to process the documents we suggest further. Our concentration on news is also novel.

\textbf{Knowledge Base Acceleration.} In this task, given specific information extraction templates, a given corpus is analyzed  in order to find worthwhile mentions of an entity or snippets that match the templates. Balog \cite{DBLP:conf/riao/BalogRTN13,DBLP:conf/sigir/BalogR13} recommend news citations for an entity. Prior to that, the news articles are classified for their appropriateness for an entity, where as features for the classification task they use entity, document, entity-document and temporal features. The best performing features are those that measure similarity between an entity and the news document. West et al.~\cite{DBLP:conf/www/WestGMSGL14} consider the problem of knowledge base completion, through question answering and complete missing facts in Freebase based on templates, i.e. \emph{Frank\_Zappa} \texttt{bornIn} \emph{Baltymore, Maryland}. 

In contrast, we do not extract facts for pre-defined templates but rather suggest news articles based on their relevance to an entity. In cases of long-tail entities, we can suggest to add a novel section  through our abstraction and generation of section templates at entity class level.

\textbf{Entity Salience.} Determining which entities are prominent or salient in a given text has a long history in NLP, sparked by the linguistic theory of Centering~\cite{walker1998centering}. Salience has been used in pronoun and co-reference resolution \cite{DBLP:conf/acl/Ng10}, or to predict which entities will be included in an abstract of an article \cite{DBLP:conf/eacl/DunietzG14}. Frequent features to measure salience include the frequency of an entity in a document, positioning of an entity, grammatical function or internal entity structure (POS tags, head nouns etc.). These approaches are not currently aimed at knowledge base generation or Wikipedia coverage extension but we postulate that an entity's salience in a news article is a prerequisite to the news article being relevant enough to be included in an entity page. We therefore use the salience features in~\cite{DBLP:conf/eacl/DunietzG14}  as part of our model.  However, these features are document-internal --- we will show that they are not sufficient to predict news inclusion into an entity page and add features of entity authority, news authority and novelty that measure the relations between several entities, between entity and news article as well as between several competing news articles.

\section{Problem Definition and \\Approach Outline}\label{sec:problem_definition}

\subsection{Terminology and Problem Definition}\label{subsec:terminology}

We are interested in named entities mentioned in documents. An entity $e$ can be identified by a canonical name, and can be \emph{mentioned} differently in text via different \emph{surface forms}. We canonicalize these mentions to entity pages in Wikipedia, a method  typically known as \emph{entity linking}. We denote the set of canonicalized entities extracted and linked from a news article $n$ as $\varphi(n)$. For example, in Figure~\ref{fig:approach}, entities are canonicalized into Wikipedia entity pages (e.g. \emph{Odisha} is canonicalized to the corresponding article\footnote{\url{http://en.wikipedia.org/wiki/Odisha}}). For a collection of news articles $\mathbf{N}$, we further denote the resulting set of entities by $\mathbf{E}=\cup_{n \in \mathbf{N}}{\{e_i\}}$. 

Information in an entity page is organized into sections and evolves with time as more content is added. We refer to the state of Wikipedia at a time $t$ as $\mathcal{W}_t$ and the set of sections for an entity page $e$ as its \emph{entity profile} $S_e(t)$. Unlike news articles, text in Wikipedia could be explicitly linked to entity pages through anchors. The set of entities explicitly referred in text from section $s \in S_e(t)$ is defined as $\gamma(s)$. Furthermore, Wikipedia induces a category structure over its entities, which is exploited by knowledge bases like YAGO (e.g. \emph{Barack\_Obama} \texttt{isA Person}). Consequently, each entity page belongs to one or more entity categories or classes $c$. Now we can define our news suggestion problem below:

\begin{definition}[News Suggestion Problem] Given a set of news articles $\mathbf{N}=\{n_1,\ldots,n_k\}$ and set of Wikipedia entity pages $\mathbf{E}=\{e_1,\ldots, e_m\}$ (from $\mathcal{W}_t$) we intend to suggest a news article $n$ published at time $t_i > t$ to entity page $e$ and additionally to the most relevant section for the entity page $s \in S_e(t)$.
\end{definition}

\subsection{Approach Overview}

We approach the news suggestion problem by decomposing it into two tasks:

\begin{enumerate}
\itemsep0em
 \item \emph{AEP}: \emph{Article--Entity} placement
 \item \emph{ASP}: \emph{Article--Section} placement
 \end{enumerate}

In this first step, for a given entity-news pair $\langle n,
e\rangle$, we determine whether the given news article
$n\in\mathbf{N}$ should be suggested (we will refer to this as
\emph{`relevant'}) to entity $e\in \mathbf{E}$. To generate such
$\langle n, e\rangle$ pairs, we perform the \emph{entity linking}
process, $\varphi(n)$, for $n$. 

The \emph{article--entity} placement task (described in detail in Section~\ref{subsec:article_linking}) for a pair $\langle n, e\rangle$ outputs a binary label (either \emph{`non-relevant'} or \emph{`relevant'}) and is formalized in Equation~\ref{eq:article_entity}.
\begin{equation}\label{eq:article_entity}
AEP: \langle e, n\rangle \rightarrow \{0,1\}, \,\,\,\; \forall e \in \varphi(n) \; \wedge\; n \in \mathbf{N}
\end{equation}

In the second step, we take into account all \emph{`relevant'} pairs $\langle n, e\rangle$ and find the correct \emph{section} for article $n$ in entity $e$, respectively its profile $S_e(t)$ (see Section~\ref{subsec:section_linking}). The \emph{article--section} placement task, determines the correct section for the triple $\langle n, e, S_e(t)\rangle$, and is formalized in Equation~\ref{eq:article_section}.
\begin{equation}\label{eq:article_section}
ASP: \langle e, n, S_e(t) \rangle \rightarrow \{s_1,\ldots, s_{k}\}, \; s \in S_e(t)
\end{equation}

In the subsequent sections we describe in details how we approach the two tasks for suggesting news articles to entity pages.

\section{News Article Suggestion}\label{sec:approach}

In this section, we provide an overview of the \emph{news suggestion} approach to Wikipedia entity pages (see Figure~\ref{fig:approach}). The approach is split into two tasks: (i) \emph{article-entity} (\emph{AEP}) and (ii) \emph{article-section} (\emph{ASP}) placement. For a Wikipedia snapshot $\mathcal{W}_t$ and a news corpus $\mathbf{N}$, we first determine which news articles should be suggested to an entity $e$.  We will denote our approach for \emph{AEP} by $\mathcal{F}_e$. Finally, we determine the most appropriate section for the \emph{ASP} task and we denote our approach with $\mathcal{F}_s$.

In the following, we describe the process of learning the functions $\mathcal{F}_e$ and $\mathcal{F}_s$. We introduce features for the learning process, which encode information regarding the entity \emph{salience}, \emph{relative authority} and \emph{novelty} in the case of AEP task. For the \emph{ASP} task, we measure the \emph{overall fit} of an article to the entity sections, with the entity being an input from \emph{AEP} task. Additionally, considering that the entity profiles $S_e(t)$ are incomplete, in the case of a missing section we suggest and expand the entity profiles based on \emph{section templates} generated from entities of the same class $c$ (see Section~\ref{subsubsec:as_sectiontemplates}).

\subsection{Article--Entity Placement}\label{subsec:article_linking}

In this step we learn the function $\mathcal{F}_e$ to correctly determine whether $n$ should be suggested for $e$, basically a binary classification model (0=\emph{`non-relevant'} and 1=\emph{`relevant'}). Note that we are mainly interested in finding the \emph{relevant} pairs in this task. For every news article, the number of disambiguated entities is around 30 (but $n$ is suggested for only two of them on average). Therefore, the distribution of \emph{`non-relevant'} and \emph{`relevant'} pairs is skewed towards the earlier, and by simply choosing the \emph{`non-relevant'} label we can achieve a high accuracy for $\mathcal{F}_e$. 
Finding the relevant pairs is therefore a considerable challenge. 

An article $n$ is suggested to $e$ by our function $\mathcal{F}_e$ if it fulfills the following properties. The entity $e$ is \emph{salient} in $n$ (a central concept), therefore ensuring that $n$ is about $e$ and that $e$ is important for $n$. Next, given the fact there might be many articles in which $e$ is \emph{salient}, we also look at the reverse property, namely whether $n$ is important for $e$. We do this by comparing the \emph{authority} of $e$ (which is a measure of popularity of an entity, such as its frequency of mention in a whole corpus) with the authority of its co-occurring entities in $\varphi(n)$, leading to a feature we call \emph{relative authority}. The intuition is that for an entity that has overall lower authority than its co-occurring entities, a news article is more easily of importance.\footnote{This is why people occurring infrequently in the news keep any press cutting mentioning them.} Finally, if the article we are about to suggest is already covered in the entity profile $S_e(t)$, we do not wish to suggest \emph{redundant} information, hence the \emph{novelty}.
Therefore,  the learning objective of $\mathcal{F}_e$ should fulfill the following properties. Table~\ref{tbl:importance_salience} shows a summary of the computed features for $\mathcal{F}_e$.
\begin{enumerate} \itemsep0em
	\item \textbf{Salience:} entity $e$ should be a \emph{salient} entity in news article $n$ 
	\item \textbf{Relative Authority:} the set of entities $e' \in \varphi(n)$ with which $e$ co-occurs should have higher \emph{authority} than $e$, making $n$ important for $e$
	\item\textbf{Novelty:} news article $n$ should provide \emph{novel} information for entity $e$ taking into account its profile $S_{e}(t-1)$ 
\end{enumerate} 
\begin{table}[ht!]
\centering\small
\scalebox{0.9}{
\begin{tabular}{p{1.5cm} p{4.5cm} p{1.5cm}}
\toprule
feature & description\\
\midrule
$\Phi(e,n)$ & the relative frequency of $e$ in news article $n$. & \multirow{2}{1.5cm}{\emph{salience}}\\
\texttt{Baseline Features} & set of features as proposed by Dunietz and Gillick~\cite{DBLP:conf/eacl/DunietzG14} & \\[3ex]
$\widehat{\Gamma}(e|\varphi(n))$ & relative authority as the score of entities that have higher authority than $e$ and that co-occur in $n$. & \multirow{2}{1.5cm}{\emph{authority}}\\
$P(D)$ & measures the news domain authority. & \\
$\mathcal{N}(n|e)$ & measures the novelty of a news article $n$ for a given entity $e$ & \emph{novelty}\\
\bottomrule
\end{tabular}}
\caption{\small{\emph{Article--Entity} placement feature summary.}}
\label{tbl:importance_salience}
\end{table}

\subsubsection{Salience-based features}

\textbf{Baseline Features.} As discussed in Section~\ref{sec:related-work}, a variety of features that measure salience of an entity in text are available from the NLP community. We reimplemented the ones in Dunietz and Gillick~\cite{DBLP:conf/eacl/DunietzG14}. This includes a variety of features, e.g. positional features, occurrence frequency and the internal POS structure of the entity and the sentence it occurs in. Table 2 in \cite{DBLP:conf/eacl/DunietzG14} gives details. 

\textbf{Relative Entity Frequency.} Although frequency of mention and positional features play some role in baseline features, their interaction is not modeled by a single feature nor do the positional features encode more than sentence position. We therefore suggest a novel feature called \emph{relative entity frequency}, $\Phi(e,n)$, that has three properties.: (i) It rewards entities for occurring throughout the text instead of only in some parts of the text, measured by the number of paragraphs it occurs in (ii) it rewards entities that occur more frequently in the opening paragraphs of an article as we model $\Phi(e,n)$ as an \emph{exponential decay} function. The decay corresponds to the positional index of the news paragraph. This is inspired by the news-specific discourse structure that tends to give short summaries of the most important facts and entities in the opening paragraphs. (iii) it compares entity frequency to the frequency of its co-occurring mentions as the weight of an entity appearing in a specific paragraph, normalized by the sum of the frequencies of other entities in $\varphi(n)$. 
\begin{equation}\label{eq:weighted_frequency}\small
\Phi(e,n) =  \frac{|p(e,n)|}{|p(n)|}\sum\limits_{p\in p(n)}\left(\frac{tf(e,p)}{\sum\limits_{e'\neq e}tf(e',p)}\right)^{\frac{1}{p}}
\end{equation}
where, $p$ represents a news paragraph from $n$, and with $p(n)$ we indicate the set of all paragraphs in $n$. The frequency of $e$ in a paragraph $p$ is denoted by $tf(e,p)$. With $|p(e,n)|$ and $|p(n)|$ we indicate the number of paragraphs in which entity $e$ occurs, and the total number of paragraphs, respectively. 

\subsubsection{Authority-based features}

\textbf{Relative Authority.} In this case, we consider the comparative relevance of the news article to the different entities occurring in it. As an example, let us consider the meeting of the Sudanese bishop \emph{Elias Taban}\footnote{\url{http://en.wikipedia.org/wiki/Elias_Taban}} with \emph{Hillary Clinton}\footnote{\url{http://en.wikipedia.org/wiki/Hillary_Clinton}}. Both entities are salient for the meeting. However, in Taban's Wikipedia page, this meeting is discussed prominently with a corresponding news reference\footnote{\url{http://tinyurl.com/mshf7j2}}, whereas in Hillary Clinton's Wikipedia page it is not reported at all. We believe this is not just an omission in Clinton's page but mirrors the fact that for the lesser known Taban the meeting is big news whereas for the more famous Clinton these kind of meetings are a regular occurrence, not all of which can be reported in what is supposed to be a selection of the most important events for her. Therefore, if two entities co-occur, the news is more relevant for the entity with the lower a priori authority.

The \emph{a priori authority} of an entity (denoted by $\Gamma(e)$) can be measured in several ways. We opt for two approaches: (i) probability of entity $e$ occurring in the corpus $\mathbf{N}$, and (ii) authority assessed through centrality measures like PageRank~\cite{page1999pagerank}. For the second case we construct the graph $G=(V,E)$ consisting of entities in $\mathbf{E}$ and news articles in $\mathbf{N}$ as \emph{vertices}. The \emph{edges} are established between $n$ and entities in $\varphi(n)$, that is $\langle n \rightarrow \varphi(n)\rangle$, and the out-links from $e$, that is $\langle e \rightarrow \gamma(s_(t-1))\rangle$ (arrows present the \emph{edge} direction).

Starting from a priori authority, we proceed to \emph{relative authority} by comparing the a priori authority of co-occurring entities in  $\varphi(n)$. We define the \emph{relative authority} of $e$ as the proportion of co-occurring entities $e'\in \varphi(n)$  that have a higher a priori  authority than $e$ (see Equation~\ref{eq:avg_authority}.
\begin{equation}\label{eq:avg_authority}\small
\hat{\Gamma}(e|\varphi(n)) = \frac{1}{|\varphi(n)|}\sum\limits_{e'\in \varphi(n)}\mathbbm{1}_{\Gamma(e') > \Gamma(e)}
\end{equation}
As we might run the danger of not suggesting any news articles for entities with very high a priori authority (such as Clinton) due to the strict inequality constraint, we can relax the constraint such that the authority of co-occurring entities is above a certain threshold.

\textbf{News Domain Authority.} The news domain authority addresses two main aspects. Firstly, if bundled together with the \emph{relative authority} feature, we can ensure that dependent on the entity authority, we suggest news from authoritative sources, hence ensuring the quality of suggested articles. The second aspect is in a news streaming scenario where multiple news domains report the same event --- ideally only articles coming from authoritative sources would fulfill the conditions for the news suggestion task.

The \emph{news domain} authority is computed based on the number of news references in Wikipedia coming from a particular \emph{news domain} $D$. This represents a simple prior that a news article $n$ is from domain $D$ in corpus $\mathbf{N}$. We extract the domains by taking the base URLs from the news article URLs.

\subsubsection{Novelty-based features} An important feature when suggesting an article $n$ to an entity $e$ is the \emph{novelty} of $n$ w.r.t the already existing entity profile $S_{e}(t-1)$. Studies~\cite{Bernstein:2005:RDS:1099554.1099733} have shown that on comparable collections to ours (TREC GOV2) the number of duplicates can go up to $17\%$.  This figure is likely higher for major events concerning highly authoritative entities on which all news media will report.

Given an entity $e$ and the already added news references $N_{t-1}=\{n_1,\ldots, n_k\}$ up to year $t-1$, the \emph{novelty} of $n_{k+1}$ at year $t$ is measured by the KL divergence between the language model of $n_{k+1}$ and articles in $N_{t-1}$. We combine this measure with the \emph{entity} overlap of $n_{k+1}$ and $n'\in N_{t-1}$. The \emph{novelty} value of $n_{k+1}$ is given by the minimal divergence value. Low scores indicate low novelty for the entity profile $S_e(t)$. 

\begin{align}\label{eq:novelty}\small \mathcal{N}(n|e) = 
\min\limits_{n'\in N_{t-1}}\left\{\lambda \cdot D_{KL}\left(\theta(n')
|| \theta(n)\right) + \vphantom{D_{KL}\left(\hat{\theta}(N) ||
\hat{\theta}(n)\right)}\right. \nonumber\\ \vphantom{\lambda \cdot
D_{KL}\left(\theta(n') || \theta(n)\right)}\left. (1-\lambda) \cdot
jaccard\left(\varphi(n'),\varphi(n)\right)\right\} \end{align} 
where $D_{KL}$ is the KL divergence of the language models ($\theta(n)$ and $\theta(n')$), whereas $\lambda$ is the mixing weight ($\lambda=\{0, \ldots, 1\}$) between the language models $D_{KL}$ and the entity overlap in $n$ and $n'$.

\subsection{Article--Section Placement}\label{subsec:section_linking}

\begin{table*}[ht!]
\centering\small
\scalebox{0.9}{
\begin{tabular}{l l l}
\toprule
feature type & feature & description\\
\midrule
  \multirow{2}{3cm}{\textbf{Topic}} & $jaccard(LDA(n), LDA(s(t-1)))$ & \multirow{2}{8cm}{Topic similarity between an article $n$ and the (entity) section text, and with already referenced news articles in a given entity section.}\\
  &  $jaccard(LDA(n), N_{t-1})$ & \\[3ex]

 \textbf{Syntactic} & \emph{POS -- sim}   & \multirow{2}{8cm}{POS tag overlap (uni/bi/trigrams) between a news article and the section text.}\\[1.5ex]
 
 \multirow{3}{3cm}{\textbf{Lexical}} & $jaccard(title(n), s(t-1))$ &  \multirow{3}{8cm}{News title and top--$k$ paragraphs ($k=1 \ldots 5$) similarity with (entity) section text.}\\
 & $D_{KL}(\theta(p(k) || \theta(s(t-1)))$ & \\
 & $cos(p(n), s(t-1))$ &  \\[1.5ex]

 \multirow{2}{3cm}{\textbf{Entity-based}} & $jaccard(\varphi(n), \gamma(s,t-1))$ & \multirow{2}{8cm}{Entity and entity class overlap between the news article and entities appearing in a specific entity section.}\\
 & $jaccard(\text{\texttt{typeOf}}(\varphi(n)), \text{\texttt{typeOf}}(\gamma(s(t-1))))$ & \\[1.5ex]
 
 \multirow{2}{3cm}{\textbf{Frequency}} & \texttt{\#POS,\#paragraphs,$|n|, |\varphi(n)|$} & \multirow{2}{8cm}{Frequency based features of the different POS tags, number of paragraphs, entities that are found in a news article}\\
 & \texttt{top--$k(e)$, top--$k(\text{\texttt{typeOf}}(e))$} & \\
 \bottomrule
\end{tabular}}
\caption{\small{Feature types used in $\mathcal{F}_s$ for suggesting news articles into the entity sections. We compute the features 
for all $s\in \widehat{S}_c(t-1)$ as well as $s_e(t-1)$}.}
\label{tbl:feature_list}
\end{table*}

We model the \emph{ASP} placement task as a successor of the \emph{AEP} task. For all the \emph{`relevant'} news entity pairs, the task is to determine the correct entity section. Each section in a Wikipedia entity page represents a different topic. For example, \emph{Barack Obama} has the sections  \emph{`Early Life', `Presidency', `Family and Personal Life'} etc. However, many entity pages have an  incomplete section structure. Incomplete or missing sections are due to two Wikipedia properties. First, long-tail entities miss information and sections due to their lack of popularity. Second, for all entities whether popular or not, certain sections might occur for the first time due to real world developments. As an example, the entity \emph{Germanwings} did not have an \emph{`Accidents'} section before this year's disaster, which was the first in the history of the airline.

Even if sections are missing for certain entities, similar sections usually occur in other entities of the same class (e.g. other airlines had disasters and therefore their pages have an accidents section). We exploit such homogeneity of section structure and construct templates that we use to expand entity profiles. The learning objective for $\mathcal{F}_s$ takes into account the following properties:
\begin{enumerate}
	\itemsep0em
	\item \textbf{Section-templates:} account for incomplete section structure for an entity profile $S_e(t)$ by constructing section templates $\widehat{S}_c$ from an entity class $c$
	\item \textbf{Overall fit:} measures the overall fit of a news article to sections in the section templates $\widehat{S}_c$
\end{enumerate}

\subsubsection{Section-Template Generation}\label{subsubsec:as_sectiontemplates}

Given the fact that \emph{entity profiles} are often incomplete, we  construct \emph{section templates} for every \emph{entity class}. We group entities based on their class $c$ and construct \emph{section templates} $\widehat{S}_c$. For different entity classes, e.g. \texttt{Person} and \texttt{Location}, the section structure and the information represented in those section varies heavily. Therefore, the section
templates are with respect to the individual classes in our experimental setup (see Figure~\ref{fig:entity_distribution}). 
\begin{equation}\label{eq:section_template}\small
\widehat{S}_c = \{s_1,\ldots, s_k\}, \forall S_e(t) \in \mathbf{E}
\wedge e \text{\texttt{ typeOf }} c \end{equation}

Generating \emph{section templates} has two main advantages. Firstly, by considering class-based profiles, we can overcome the problem of incomplete individual entity profiles and thereby are able to suggest news articles to sections that do not yet exist in a specific entity $S_e(t)$. The second advantage is that we are able to canonicalize the sections, i.e. \emph{`Early Life'} and \emph{`Early Life and Childhood'} would be treated similarly.

To generate the section template $\widehat{S}_c$, we extract all sections from entities of a given type $c$ at year $t$. Next, we cluster the entity sections, based on an extended version of \emph{k--means} clustering \cite{DBLP:journals/pami/KanungoMNPSW02}, namely \emph{x--means} clustering introduced in Pelleg et al. which estimates the number of clusters efficiently \cite{pelleg2000x}. As a similarity metric we use the cosine similarity computed based on the \emph{tf--idf} models of the sections. Using the \emph{x--means} algorithm we overcome the requirement to provide the number of clusters \emph{k} beforehand. \emph{x--means} extends the \emph{k--means} algorithm, such that a user only specifies a range [$K_{min}$, $K_{max}$] that the number of clusters may reasonably lie in.

\subsubsection{News-section fit}

The learning objective of $\mathcal{F}_s$ is to determine the overall fit of a news article $n$ to one of the sections in a given section template $\widehat{S}_c$. The template is pre-determined by the class of the entity for which the news is suggested as relevant by $\mathcal{F}_e$. In all cases, we measure how well $n$ fits each of the sections $s\in \widehat{S}_{c}(t-1)$ as well as the specific entity section $s' \in S_e(t-1)$. The section profiles in $\widehat{S}_c(t-1)$ represent the aggregated entity profiles from all entities of class $c$ at year $t-1$.

To learn $\mathcal{F}_s$ we rely on a variety of features that consider several similarity aspects as shown in Table~\ref{tbl:feature_list}. For the sake of simplicity we do not make the distinction in Table~\ref{tbl:feature_list} between the individual entity section and class-based section similarities, $s_{e}(t-1)$ and $s(t-1)$, respectively. Bear in mind that an entity section $s_e$ might be present at year $t$ but not at year $t-1$ (see for more details the discussion on entity profile expansion in Section~\ref{subsubsec:profile_expansion}).

\textbf{Topic.} We use topic similarities to ensure (i) that the content of $n$ fits topic-wise with a specific section text and (ii) that it has a similar topic to previously referred news articles in
that section. In a pre-processing stage we compute the topic models for the news articles, entity sections $S_{e}(t-1)$ and the aggregated class-based sections in $\widehat{S}_{c}$. The topic models are computed using LDA~\cite{Blei:2003:LDA:944919.944937}. We only computed a single topic per article/section as we are only interested in topic term overlaps between article and sections. We distinguish two main features: the first feature measures the overlap of topic terms between $n$ and the entity section $s_{e}(t-1)$ and $s(t-1) \in \widehat{S}_c$, and the second feature measures the overlap of the topic model of $n$ against referred news articles in $N_{t-1}$ at time $t-1$.

\textbf{Syntactic.} These features represent a mechanism for conveying the importance of a specific text snippet, solely based on the frequency of specific POS tags (i.e. \texttt{NNP, CD} etc.), as commonly used in text summarization tasks. Following the same intuition as in \cite{DBLP:conf/acl/SauperB09}, we weigh the importance of articles by the count of specific POS tags. We expect that for different sections, the importance of POS tags will vary. We measure the similarity of POS tags in a news article against the section text. Additionally, we consider \emph{bi-gram} and \emph{tri-gram} POS tag overlap. This exploits similarity in syntactical patterns between the news and section text.

\textbf{Lexical.} As \emph{lexical} features, we measure the similarity of $n$ against the entity section text $s_{e}(t-1)$ and the aggregate section text $s(t-1)$. Further, we distinguish between the
overall similarity of $n$ and that of the different news paragraphs ($p(n)$ which denotes the paragraphs of $n$ up to the 5th paragraph). A higher similarity on the first paragraphs represents a more confident indicator that $n$ should be suggested to a specific section $s$. We measure the similarity based on two metrics: (i) the KL-divergence between the computed \emph{language models} and (ii) \emph{cosine} similarity of the corresponding paragraph text $p(n)$ and section text.

\textbf{Entity-based.} Another feature set we consider is the overlap of \emph{named entities} and their corresponding \emph{entity classes}. For different entity sections, we expect to find a particular set of entity classes that will correlate with the section, e.g. `\emph{Early Life}' contains mostly entities related to family, school, universities etc.

\textbf{Frequency.} Finally, we gather statistics about the number of entities, paragraphs, news article length, top--$k$ entities and entity classes, and the frequency of different POS tags. Here we try to capture patterns of articles that are usually cited in specific sections.

\section{Datasets and Pre-Processing}\label{sec:datasets}

\subsection{Evaluation Plan}
In this section we outline the evaluation plan to verify the effectiveness of our learning approaches. To evaluate the news suggestion problem we are faced with two challenges. 

\begin{itemize}
	\itemsep0em
	\item \emph{What comprises the ground truth for such a task ?}
	\item \emph{How do we construct training and test splits given that entity pages consists of  text added at different points in time ?}
\end{itemize}

Consider the ground truth challenge. Evaluating if an arbitrary news article should be included in Wikipedia is both subjective and difficult for a human if she is not an expert. An invasive approach,
which was proposed by Barzilay and Sauper~\cite{DBLP:conf/acl/SauperB09}, adds content directly to Wikipedia and expects the editors or other users to redact irrelevant content over a period of time. The limitations of such an evaluation technique is that content added to long-tail entities might not be evaluated by informed users or editors in the experiment time frame. It is hard to estimate how much time the added content should be left on the entity page. A more non-invasive approach could involve crowdsourcing of entity and news article pairs in an IR style relevance assessment setup. The problem of such an approach is again  finding knowledgeable users or experts for long-tail entities. Thus the notion of \emph{relevance} of a news recommendation is challenging to evaluate in a crowd setup.

We take a slightly different approach by making an assumption that the news articles already present in Wikipedia entity pages are relevant. To this extent, we extract a dataset comprising of all news articles referenced in entity pages (details in Section~\ref{subsec:datasets}). At the expense of not evaluating the space comprising of  news articles absent in Wikipedia, we succeed in (i) avoiding restrictive assumptions about the quality of human judgments, (ii) being invasive and polluting Wikipedia, and (iii) deriving a reusable test bed for quicker experimentation.

The second challenge of construction of training and test set separation is  slightly easier and is addressed in Section~\ref{subsec:test_train}.

\subsection{Datasets}
\label{subsec:datasets}

\begin{figure}[h!]
\includegraphics[width=0.9\columnwidth]{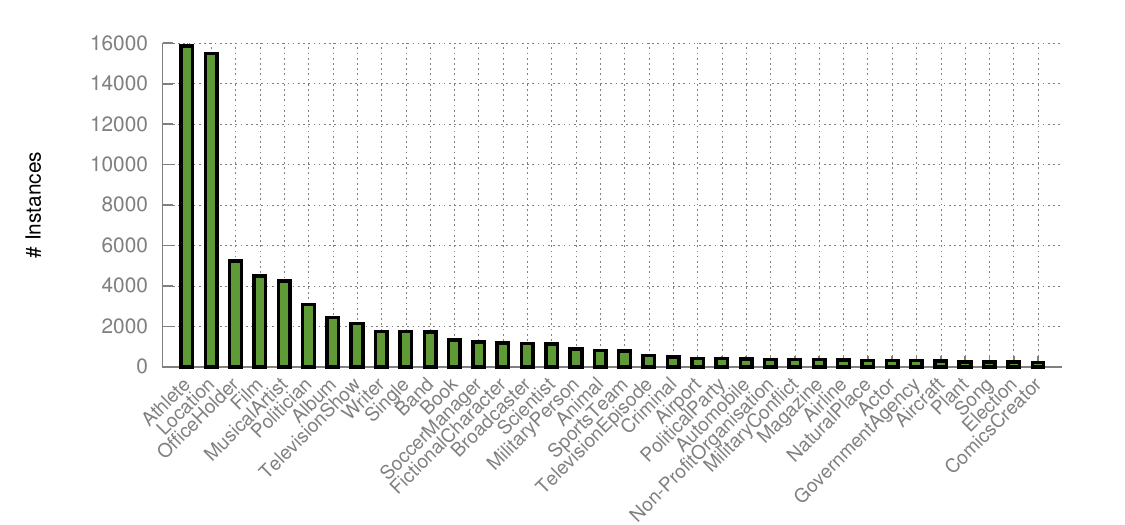}
\caption{\small{Number of entities with at least one news reference for different entity classes.}}
\label{fig:entity_distribution}
\end{figure}

The datasets we use for our experimental evaluation are directly extracted from the Wikipedia entity pages and their  revision history. The generated data represents one of the contributions of our paper.\footnote{\url{http://l3s.de/~fetahu/cikm2015/data/}} The datasets are the following:

\textbf{Entity Classes.} We focus on a manually predetermined set of \emph{entity classes} for which we expect to have news coverage. The number of analyzed \emph{entity classes} is $27$, including $73,734$
entities with at least one news reference. The \emph{entity classes} were selected from the DBpedia class ontology. Figure~\ref{fig:entity_distribution} shows the number of entities per class for the years (2009-2014).

\textbf{News Articles.} We extract all news references from the collected Wikipedia entity pages.\footnote{A news reference in Wikipedia is denoted by the template \texttt{\{cite type=`news' | url=`'\}}} The extracted news references are associated with the sections in which they appear. In total there were $411,673$ news references, and after crawling we end up with $351,982$ successfully crawled news articles. The details of the news article distribution, and the number of entities and sections from which they are referred are shown in Table~\ref{tbl:news_dist}. 

\begin{table}[h!]\small
\centering
\scalebox{0.9}{
\begin{tabular}{c c c c }
\toprule
\texttt{year} & \texttt{\#news} & \texttt{\#entities} & \texttt{\#sections}\\
\midrule
\texttt{2009} & 42707 & 13550 & 3510 \\
\texttt{2010} & 78328 & 24953 & 8416 \\
\texttt{2011} & 73491 & 23144 & 6581 \\
\texttt{2012} & 81473 & 25980 & 8455 \\
\texttt{2013} & 69079 & 22121 & 8183 \\
\texttt{2014} & 29961 & 11088 & 4694 \\
\bottomrule
\end{tabular}}
\caption{\small{News articles, entities and sections distribution across years.}}
\label{tbl:news_dist}
\end{table}

\textbf{Article-Entity Ground-truth.} The dataset comprises of the news and entity pairs $\langle n, e\rangle \rightarrow \{0,1\}$. News-entity pairs are relevant if the news article is referenced in the entity page. Non-relevant pairs (i.e. negative training examples) consist of news articles that contain an entity but are not referenced in that entity's page. If a news article $n$ is referred from $e$ at year $t$, the features are computed taking into account the entity profiles at year $S_e(t-1)$.

\textbf{Article-Section Ground-truth.} The dataset consists of the triple $\langle n, e, s\rangle$, where $s \in \widehat{S}_c$, where we assume that  $\langle n, e\rangle$ has already been determined as relevant.  We therefore have a multi-class classification problem where we need to determine the section of $e$ where $n$ is cited. Similar to the \emph{article-entity} ground truth, here too the features compute the similarity between $n$, $S_e(t-1)$ and $\widehat{S}_c(t-1)$.

\subsection{Data Pre-Processing}\label{subsec:pre_processing} 

We POS-tag the news articles and entity profiles $S_e(t)$ with the Stanford tagger~\cite{Toutanova:2003:FPT:1073445.1073478}.  For entity linking the news articles, we use TagMe!\cite{DBLP:journals/software/FerraginaS12} with a confidence score of 0.3. On a manual inspection of a random sample of 1000 disambiguated entities, the accuracy is above 0.9.  On average, the number of entities per news article is approximately 30. For entity linking the entity profiles, we simply follow the \emph{anchor} text that refers to Wikipedia entities.

\subsection{Train and Testing Evaluation Setup}\label{subsec:test_train} 

We evaluate the generated supervised models for the two tasks, \emph{AEP} and \emph{ASP}, by splitting the train and testing instances. It is important to note that for the pairs $\langle n, e\rangle$ and the triple $\langle n,e,\widehat{S}_{c}\rangle$, the news article $n$ is referenced at time $t$ by entity $e$, while the features take into account the entity profile at time $t-1$. This avoids any `overlapping' content between the news article and the entity page, which could affect the learning task of the functions $\mathcal{F}_e$ and $\mathcal{F}_s$. Table~\ref{tbl:train_test_instances} shows the statistics of train and test instances. We learn the functions at year $t$ and test on instances for the years greater than $t$. Please note that we do not show the performance for year 2014 as we do not have data for 2015 for evaluation.

\begin{table}[h!]
\centering\small
\scalebox{0.9}{
\begin{tabular}{p{1cm} l p{1.5cm} l l}
\toprule
 & \multicolumn{2}{c}{$\mathcal{F}_e$} & \multicolumn{2}{c}{$\mathcal{F}_s$}\\
 \toprule
 & \texttt{train} & \texttt{test} & \texttt{train} & \texttt{test} \\
\midrule
2009 & 74,005 & 469,386 & 19,399 & 218,757\\
2010 & 190,409 & 382,085 & 70,486 & 167,670\\
2011 & 286,588 & 292,398 & 115,286 & 122,870\\
2012 & 386,647 & 177,755 & 170,682 & 67,474\\
2013 & 471,209 & 59,172 & 218,538 & 19,618\\
\bottomrule
\end{tabular}}
\caption{\small{Number of instances for train and test in the \emph{AEP} and \emph{ASP} tasks.}}
\label{tbl:train_test_instances}
\end{table}

\section{Results and Discussion}\label{sec:results}

\subsection{Article--Entity
Placement}\label{subsec:article_entity_results} Here we introduce the evaluation setup and analyze the results for the \emph{article--entity (AEP)} placement task. We only report  the evaluation metrics for the \emph{`relevant'} news-entity pairs. A detailed explanation on why we focus on the \emph{`relevant'} pairs is provided in
Section~\ref{subsec:article_linking}.

\subsubsection{Evaluation Setup}\label{subsubsec:article_entity_setup}
\textbf{Baselines.} We consider the following baselines for this task.

\begin{itemize}
\itemsep0em
\item \textbf{B1.} The first baseline uses only the salience-based features  by Dunietz and Gillick~\cite{DBLP:conf/eacl/DunietzG14}. 
\item \textbf{B2.} The second baseline assigns the value \emph{relevant} to a pair $\langle n, e\rangle$, if and only if  $e$ appears in the title of $n$.
\end{itemize}

\textbf{Learning Models.} We use \emph{Random Forests}
(RF)~\cite{Breiman2001}.\footnote{Our emphasis in this paper is not a comparison of learning models but of course
other classifiers can be used for this task.}
We learn the RF on all computed features in Table~\ref{tbl:importance_salience}. The optimization on RF is done by splitting the feature space into multiple trees that are considered as ensemble classifiers. Consequently, for each classifier it computes the margin function as a measure of the average count of predicting the correct class in contrast to any other class. The higher the margin score the more robust the model.

\textbf{Metrics.} We compute \emph{precision} P, \emph{recall} R and F1 score for the {\em relevant} class. For example, precision is the number of news-entity pairs we correctly labeled as relevant compared to our ground truth divided by the number of all news-entity pairs we labeled as relevant.

\subsubsection{Approach Effectiveness}\label{subsubsec:ae_results}
The following results measure the effectiveness of our approach in three main aspects: (i) overall \emph{performance} of $\mathcal{F}_e$ and comparison to baselines, (ii) \emph{robustness} across the years, and (iii) \emph{optimal} model for the \emph{AEP} placement task.

\begin{figure*}[ht!]
\centering
        \begin{subfigure}[b]{0.37\textwidth}
                \includegraphics[width=\textwidth]{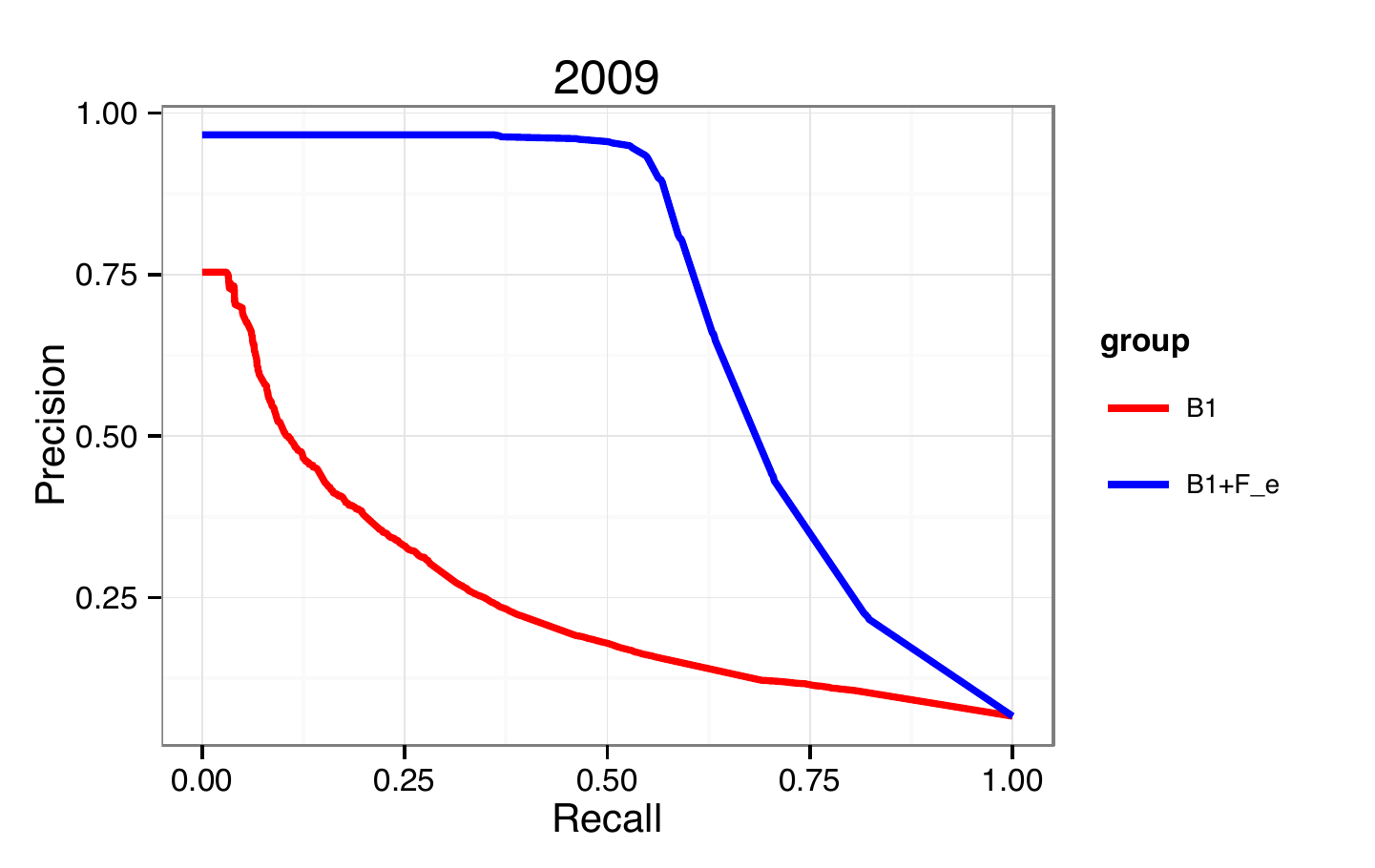}
                \caption{2009}
                \label{fig:salience_2009}
        \end{subfigure}
        \begin{subfigure}[b]{0.37\textwidth}
                \includegraphics[width=\textwidth]{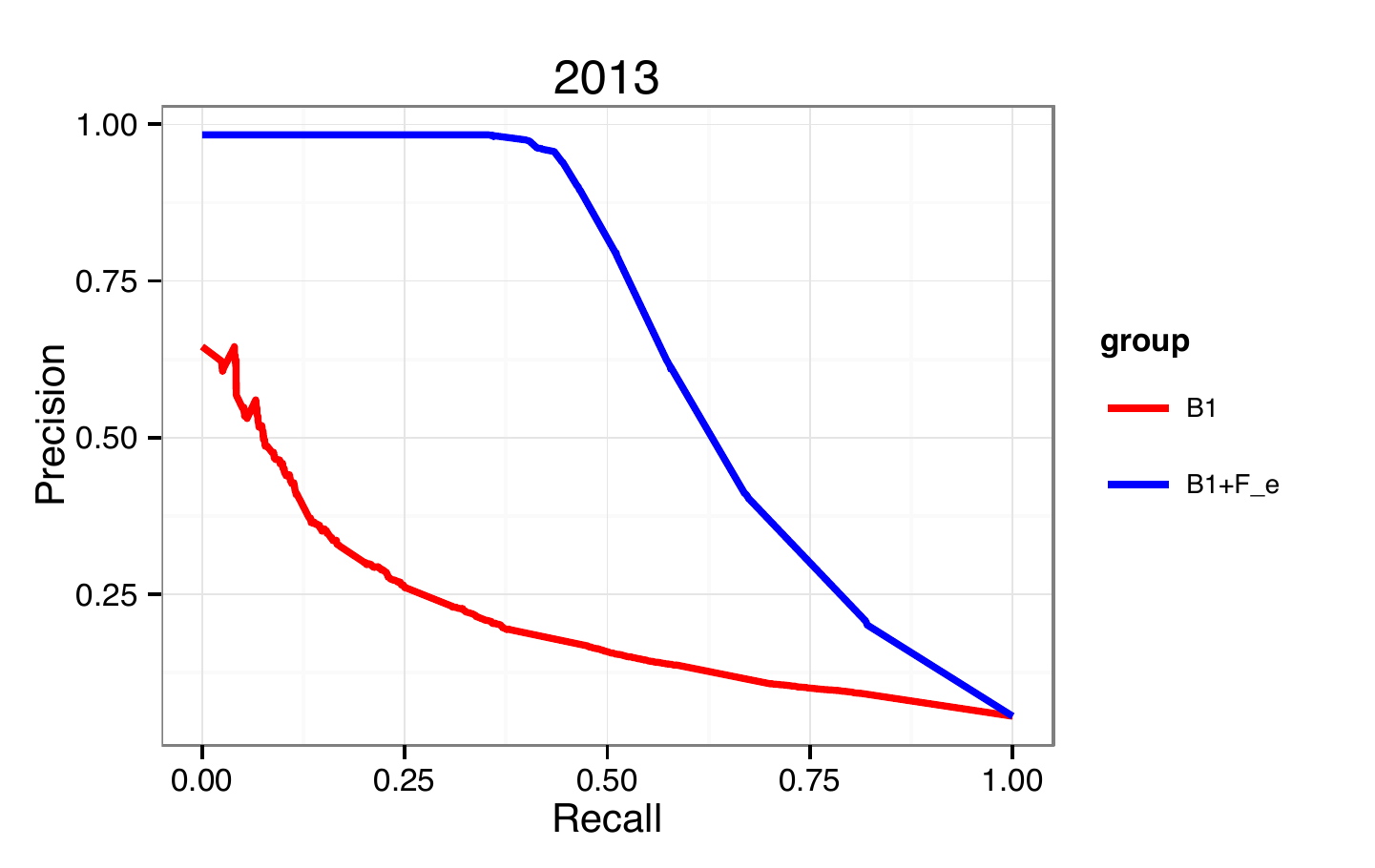}
                \caption{2013}
                \label{fig:salience_2013}
        \end{subfigure}
 	\caption{\small{Precision-Recall curve for the \emph{article--entity} placement task, in \emph{blue} is shown $\mathcal{F}_e$, and in \emph{red} is the baseline \textbf{B1}.}}
     \label{fig:salience_pr_curve}
 \end{figure*}

\textbf{Performance.} Figure~\ref{fig:salience_pr_curve} shows the results for the years  2009 and 2013, where we optimized the learning objective with instances from year $t$ and evaluate on the years $t_i>t$ (see Section~\ref{subsec:test_train}).\footnote{We  only show  the first year 2009 and the last year 2013, since the difference to the other years is marginal.}
 The results show the \emph{precision--recall} curve. The \emph{red} curve shows baseline \textbf{B1}~\cite{DBLP:conf/eacl/DunietzG14}, and the \emph{blue} one shows the performance of $\mathcal{F}_e$. The curve shows for varying \emph{confidence scores} (high to low) the precision on labeling the pair $\langle e, n\rangle$ as \emph{`relevant'}. In addition, at each \emph{confidence score} we can compute the corresponding recall for the \emph{`relevant'} label. For high confidence scores on labeling the news-entity pairs, the baseline \textbf{B1} achieves on average a precision score of P=0.50, while $\mathcal{F}_e$ has P=0.93. We note that with the drop in the confidence score the corresponding precision and recall values drop
too, and the overall F1 score for \textbf{B1} is around F1=0.2, in contrast we achieve an average score of F1=0.67.

It is evident from Figure~\ref{fig:salience_pr_curve} that for the years 2009 and 2013, $\mathcal{F}_e$ significantly outperforms  the baseline \textbf{B1}. We measure the significance through the \emph{t-test} statistic and get a \emph{p-value} of $2.2e-16$. The improvement we achieve over \textbf{B1} in absolute numbers, $\Delta$P=+0.5 in terms of precision for the years between 2009 and 2014, and a similar improvement in terms of F1 score. The improvement for recall is $\Delta$ R=+0.4. The relative improvement over \textbf{B1} for P and F1 is almost 1.8 times better, while for recall we are 3.5 times better. In Table~\ref{tbl:article_entity_results} we show the overall scores for the evaluation metrics for \textbf{B1} and
$\mathcal{F}_e$. Finally, for \textbf{B2} we achieve much poorer performance, with average scores of P=0.21, R=0.20 and F1=0.21.

\textbf{Robustness.} In Table~\ref{tbl:article_entity_results}, we show the overall performance for the years between 2009 and 2013. An interesting observation we make is that we have a very robust performance and the results are stable across the years. If we consider the experimental setup, where for year $t=2009$ we optimize the learning objective with only 74k training instances and evaluate on the rest of the instances, it achieves a very good performance. We predict with F1=0.68 the remaining 469k instances for the years $t\in (2009, 2014]$.

The results are particularly promising considering the fact that the distribution between our two classes is highly skewed.  On average the number of \emph{`relevant'} pairs account for only around $4-6\%$ of all pairs. A good indicator to support such a statement is the \emph{kappa} (denoted by $\kappa$) statistic. $\kappa$ measures agreement between the algorithm and the gold standard on both labels while correcting for chance agreement (often expected due to extreme distributions). The $\kappa$ scores for \textbf{B1} across the years is on average $0.19$, while for $\mathcal{F}_e$ we achieve a score of $0.65$ (the maximum score for $\kappa$ is 1).

\begin{table}[h!]\small
\centering
\scalebox{0.9}{
\begin{tabular}{p{1cm} l p{1cm} l p{1cm} l l}
\toprule
\texttt{year} & \multicolumn{2}{c}{P} & \multicolumn{2}{c}{R} & \multicolumn{2}{c}{F1}\\
\midrule
& \textbf{B1} & $\mathcal{F}_e$ & \textbf{B1} & $\mathcal{F}_e$ & \textbf{B1} & $\mathcal{F}_e$ \\
\midrule
2009 & 0.450 & 0.930 & 0.143 & 0.550 & 0.216 & 0.691\\
2010 & 0.503 & 0.939 & 0.128 & 0.540 & 0.204 & 0.685\\
2011 & 0.475 & 0.937 & 0.133 & 0.520 & 0.208 & 0.669\\
2012 & 0.476 & 0.935 & 0.110 & 0.515 & 0.177 & 0.664\\
2013 & 0.407 & 0.939 & 0.116 & 0.445 & 0.181 & 0.674\\
\bottomrule
\end{tabular}}
\caption{\emph{Article--Entity} placement task performance.}
\label{tbl:article_entity_results}
\end{table}

\subsubsection{Feature Analysis}\label{subsubsec:ae_features} 

\begin{figure}[ht!]
\centering
    \includegraphics[width=0.8\columnwidth]{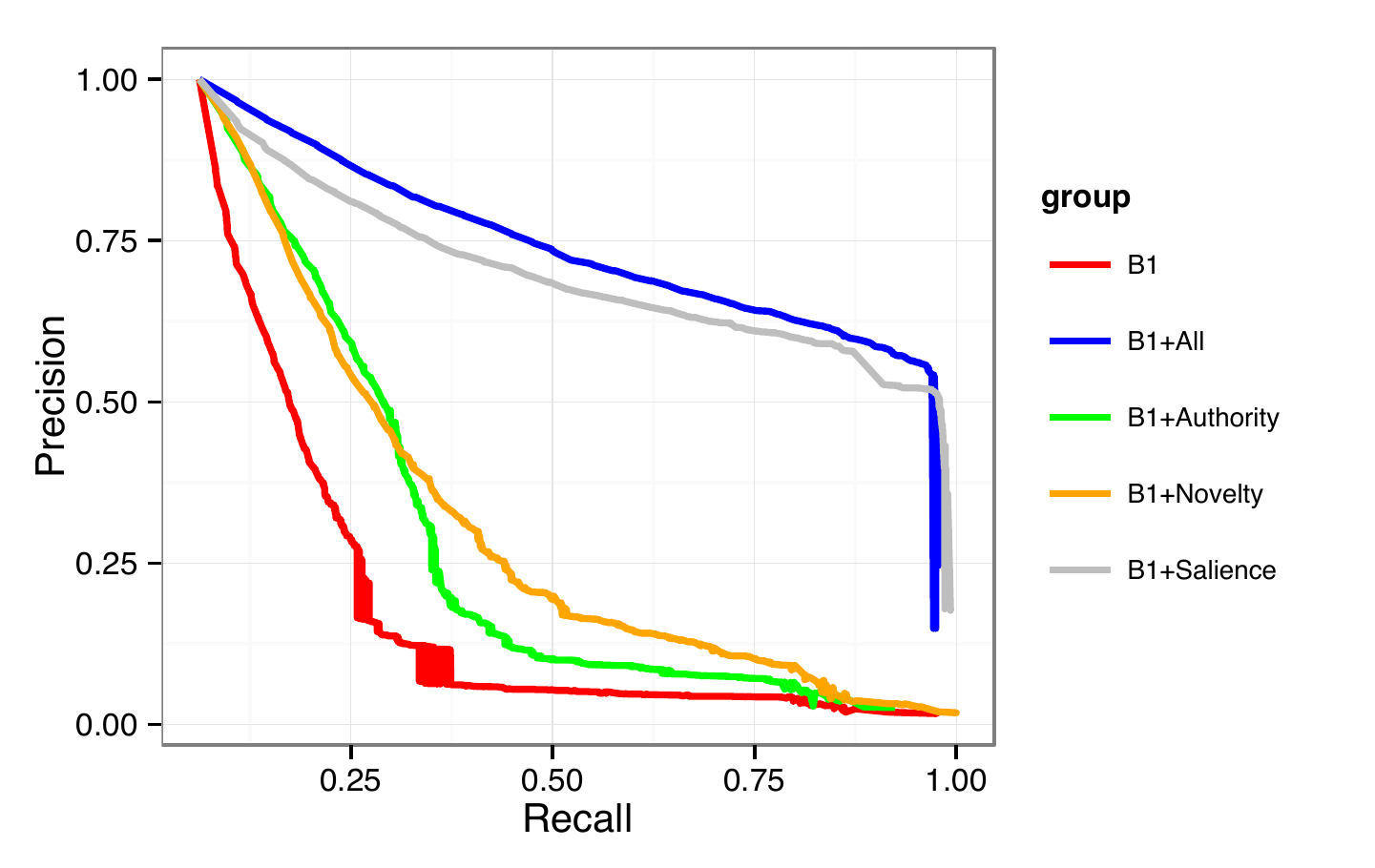}
    \caption{\small{Feature analysis for the \emph{AEP} placement task for $t=2009$.}}\
    \label{fig:salience_feature_pr_curve}
\end{figure}

In Figure~\ref{fig:salience_feature_pr_curve} we show the impact of the individual feature groups that contribute to the superior performance in comparison to the baselines. \emph{Relative entity frequency} from the \emph{salience} feature, models the entity salience as an exponentially decaying function based on the positional index of the paragraph where the entity appears. The performance of $\mathcal{F}_e$ with \emph{relative entity frequency} from the \emph{salience} feature group is close to that of all the features combined. The \emph{authority} and \emph{novelty} features account to a further improvement in terms of precision, by adding roughly a 7\%-10\% increase. However, if both feature groups are considered separately, they significantly outperform the baseline \textbf{B1}.

\subsection{Article-Section Placement}\label{subsec:article_section_results}
Here we show the evaluation setup for \emph{ASP} task and discuss the results with a focus on three main aspects, (i) the overall performance across the years, (ii) the \emph{entity class} specific performance, and (iii) the impact on \emph{entity profile} expansion by suggesting missing sections to entities based on the pre-computed templates.

\subsubsection{Evaluation Setup}\label{subsubsubsec:as_baselines}

\textbf{Baselines.} To the best of our knowledge, we are not aware of any comparable approach for this task. Therefore, the baselines we consider are the following:
\begin{itemize}
    \itemsep0em
    \item \textbf{S1}: Pick the section from template $\widehat{S}_c$ with the highest lexical similarity to $n$: \textbf{S1}$=\argmax_{s \in \widehat{S}_c(t-1)} \langle n, e, s\rangle$
    \item \textbf{S2}: Place the news into the most frequent section in $\widehat{S}_c$
\end{itemize}

\textbf{Learning Models.} We use  \emph{Random Forests} (RF)~\cite{Breiman2001} and \emph{Support Vector Machines} (SVM)~\cite{chang2011libsvm}. The models are optimized taking into account the features in Table~\ref{tbl:feature_list}. In contrast to the \emph{AEP} task, here the scale of the number of instances allows us to learn the SVM models. The SVM model is optimized using the $\epsilon-SVR$ \emph{loss} function and uses the \emph{Gaussian} kernels. 

\textbf{Metrics.} We compute \emph{precision} P as the ratio of news
for which we pick a section $s$ from $\widehat{S}_c$ and $s$ conforms
to the one in our ground-truth (see
Section~\ref{subsec:datasets}). The definition of \emph{recall} R and
F1 score follows from that of precision.

\subsubsection{Overall Article-Section Performance}\label{subsubsec:as_results}

\begin{figure}[h!]
    \centering
    \includegraphics[width=0.8\columnwidth]{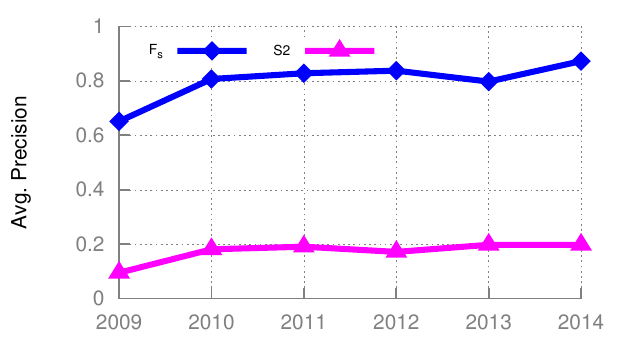}
    \caption{\small{\emph{Article-Section} performance averaged for all entity classes for $\mathcal{F}_s$ (using SVM) and \textbf{S2}.}}
     \label{fig:incremental_learning}
 \end{figure}

Figure~\ref{fig:incremental_learning} shows the overall performance and a comparison of our approach (when $\mathcal{F}_s$ is optimized using SVM) against the best performing baseline \textbf{S2}. With the
increase in the number of training instances for the \emph{ASP} task the performance is a monotonically non-decreasing function. For the year 2009, we optimize the learning objective of $\mathcal{F}_s$ with
around 8\% of the total instances, and evaluate on the rest. The performance on average is around P=0.66 across all classes. Even though for many classes the performance is already stable (as we will
see in the next section), for some classes we  improve further. If we take into account the years between 2010 and 2012, we have an increase of $\Delta$P=0.17, with around 70\% of instances used for training and
the remainder for evaluation. For the remaining years the total improvement is $\Delta$P=0.18 in contrast to the performance at year 2009.

On the other hand, the baseline \textbf{S1} has an average precision of P=0.12. The performance across the years varies slightly, with the year 2011 having the highest average precision of P=0.13. Always picking the most frequent section as in  \textbf{S2}, as shown in Figure~\ref{fig:incremental_learning}, results in an  average precision of P=0.17, with  a uniform distribution across the years.

\subsubsection{Article-Section Performance per Entity Class}

Here we show the performance of $\mathcal{F}_s$ decomposed for the different entity classes.  Specifically we analyze the 27 classes in Figure~\ref{fig:entity_distribution}. In Table~\ref{tbl:section_classifier}, we show the results for a range of years (we omit showing all years due to space constraints). For illustration purposes only, we group  them into four main classes ($\{$ \texttt{Person, Organization, Location, Event}$\}$) and into the specific sub-classes shown in the second column in Table~\ref{tbl:section_classifier}. For instance, the entity classes \texttt{OfficeHolder} and \texttt{Politician} are aggregated into \texttt{Person}--\texttt{Politics}.

It is evident that in the first year the performance is lower in contrast to the later years. This is due to the fact that as we proceed, we can better generalize and accurately determine the correct \emph{fit}
of an article $n$ into one of the sections from the pre-computed \emph{templates} $\widehat{S}_c$. The results are already stable for the year range $(2009, 2012]$. For a few \texttt{Person} sub-classes,
e.g. \texttt{Politics}, \texttt{Entertainment}, we achieve an F1 score above 0.9. These additionally represent classes with a sufficient number of training instances for the years $[2009, 2012]$. The lowest
F1 score is for the \texttt{Criminal} and \texttt{Television} classes.  However, this is directly correlated with the insufficient number of instances.

The baseline approaches for the \emph{ASP} task perform poorly. \textbf{S1}, based on  \emph{lexical similarity}, has a varying performance for different entity classes. The best performance is achieved for the class \texttt{Person -- Politics}, with P=0.43. This highlights the importance of our feature choice and that the \emph{ASP} cannot be considered as a \emph{linear function}, where the maximum similarity yields the best results. For different entity classes different features and combination of features is necessary. Considering that \textbf{S2} is the overall best performing baseline, through our approach $\mathcal{F}_s$ we have a significant improvement of over $\Delta$P=+0.64.

The models we learn are very robust and obtain high accuracy, fulfilling our pre-condition for accurate news suggestions into the entity sections. We measure the robustness of $\mathcal{F}_s$ through the $\kappa$ statistic. In this case, we have a model with roughly 10 labels (corresponding to the number of sections in a template $\widehat{S}_c$). The score we achieve shows that our model predicts with high confidence with $\kappa=0.64$.

\begin{table*}[ht!]\small
\centering
\scalebox{0.9}{
\begin{tabular}{l p{2.5cm} l l p{1.5cm} l l p{1.5cm} l l l}
\toprule
\texttt{Entity class} & \texttt{Sub-Class} & \multicolumn{3}{c}{\texttt{2009}} & \multicolumn{3}{c}{\texttt{(2009,2012]}} & \multicolumn{3}{c}{\texttt{(2012,2014]}}\\
\midrule
& & P & R & F1 & P & R & F1 & P & R & F1 \\
\midrule

\multirow{6}{*}{\texttt{Person}} &  \texttt{Entertainment} & 0.737 & 0.815 & 0.764 & 0.912 & 0.941 & 0.923 & 0.963 & 0.976 & 0.969\\
&  \texttt{Politics} & 0.916 & 0.943 & 0.930  & 0.923 & 0.948 & 0.933 & 0.936 & 0.958 & 0.946\\
&  \texttt{Scientists} & 0.467 & 0.681 & 0.554 & 0.890 & 0.940 & 0.914 & 0.931 & 0.951 & 0.938\\
&  \texttt{Sports} & 0.820 & 0.872 & 0.836 & 0.868 & 0.912 & 0.885 & 0.929 & 0.955 & 0.941\\
&  \texttt{Military} & 0.688 & 0.779 & 0.721 & 0.842 & 0.908 & 0.871 & 0.882 & 0.928 & 0.903\\
&  \texttt{Criminal} & 0.647 & 0.764 & 0.682 & 0.758 & 0.704 & 0.698 & 0.693 & 0.816 & 0.743\\[1.5ex]

\texttt{Organization} & \texttt{Organization} &  0.567 & 0.649 & 0.586  & 0.794 & 0.855 & 0.817 & 0.832 & 0.869 & 0.843\\[1.5ex]

\multirow{3}{*}{\texttt{Creative Work}} & \texttt{Television} & 0.528 & 0.650 & 0.563 & 0.745 & 0.732 & 0.709 & 0.732 & 0.772 & 0.745\\
& \texttt{Music} & 0.598 & 0.620 & 0.591 & 0.860 & 0.748 & 0.762 & 0.897 & 0.936 & 0.914\\
& \texttt{Written Work} & 0.657 & 0.765 & 0.695 & 0.733 & 0.829 & 0.772 & 0.722 & 0.791 & 0.743\\[1.5ex]

\texttt{Location} &\texttt{Location} & 0.781 & 0.763 & 0.715 & 0.857 & 0.898 & 0.872 & 0.922 & 0.956 & 0.938\\[1.5ex]
\texttt{Event} &\texttt{Event} & 0.560 & 0.682 & 0.611 & 0.858 & 0.865 & 0.853 & 0.693 & 0.716 & 0.694\\
\hline\hline
& \texttt{average} & 0.663 & 0.748 & 0.687 & 0.836 & 0.856 & 0.834 & 0.844 & 0.885 & 0.860\\
\bottomrule
\end{tabular}}
\caption{\small{\emph{Article-Section} placement performance (with $\mathcal{F}_s$ learned through SVM) for the different \emph{entity classes}. The results show the standard \textbf{P/R/F1}.}}
\label{tbl:section_classifier}
\end{table*}

\subsubsection{Entity Profile Expansion}\label{subsubsec:profile_expansion}

\begin{figure}[ht!]
\centering
\includegraphics[width=0.8\columnwidth]{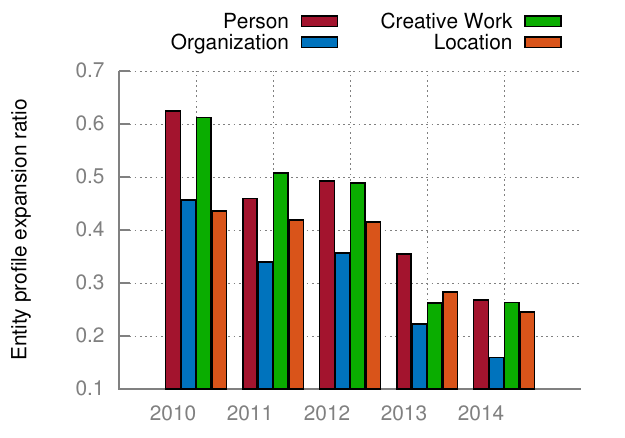}
\caption{\small{Correctly suggested news articles for $s\in {S}_e(t) \wedge s\notin {S}_e(t-1)$.}}
\label{fig:missing_sections}
\end{figure}

The last analysis is the impact we have on \emph{expanding} entity profiles $S_e(t)$ with new sections. Figure~\ref{fig:missing_sections} shows the ratio of sections for which we correctly suggest an article
$n$ to the right section in the section template $\widehat{S}_c(t)$. The ratio here corresponds to sections that are not present in the entity profile at year $t-1$, that is $s \notin S_{e}(t-1)$. However, given the generated templates $\widehat{S}_c(t-1)$, we can expand the entity profile $S_e(t-1)$ with a new section at time $t$. In details, in the absence of a section at time $t$, our model trains well on similar sections from the section template $\widehat{S}_c(t-1)$, hence we can predict accurately the section and in this case suggest its addition to the entity profile. With time, it is obvious that the expansion rate decreases at later years as the entity profiles become more `complete'. 

This is particularly interesting for expanding the entity profiles of long-tail entities as well as updating entities with real-world emerging events that are added constantly. In many cases such missing
sections are present at one of the entities of the respective entity class $c$. An obvious case is the example taken in Section~\ref{subsec:article_linking}, where the \emph{`Accidents'} is rather common for entities of type \texttt{Airline}. However, it is non-existent for some specific entity instances, i.e \emph{Germanwings} airline.

Through our \emph{ASP} approach $\mathcal{F}_s$, we are able to expand both \emph{long-tail} and \emph{trunk} entities. We distinguish between the two types of entities by simply measuring their section text length. The real distribution in the ground truth (see Section~\ref{subsec:datasets}) is 27\% and 73\% are \emph{long-tail} and \emph{trunk} entities, respectively. We are able to expand the entity profiles for both cases and all entity classes without a significant difference, with the only exception being the class \texttt{Creative Work}, where we expand significantly more \emph{trunk} entities.

\section{Conclusion and Future Work}

In this work, we have proposed an automated approach for the novel task of suggesting news articles to Wikipedia entity pages to facilitate Wikipedia updating. The process consists of two stages. In the first stage, \emph{article--entity} placement, we suggest news articles to entity pages by considering three main factors, such as \emph{entity salience} in a news article, \emph{relative authority} and \emph{novelty} of news articles for an entity page. In the second stage, \emph{article--section} placement, we determine the best fitting section in an entity page. Here, we remedy the problem of incomplete entity section profiles by constructing section templates for specific entity classes. This allows us to add missing sections to entity pages. We carry out an extensive experimental evaluation on 351,983 news articles and 73,734 entities coming from 27 distinct entity classes. For the first stage, we achieve an overall performance with P=0.93, R=0.514 and F1=0.676, outperforming our baseline competitors significantly. For the second stage, we show that we can learn incrementally to determine the correct section for a news article based on section templates. The overall performance across different classes is P=0.844, R=0.885 and F1=0.860.

In the future, we will enhance our work by extracting facts from the suggested news articles. Results suggest that the news content cited in entity pages comes from the first paragraphs. However, challenging
task such as the canonicalization and chronological ordering of facts, still remain.

\paragraph*{\textbf{Acknowledgements}} This work is funded by the ERC Advanced Grant ALEXANDRIA (grant no. 339233).

\end{document}